\documentclass{elsart}
\usepackage{graphicx}
\usepackage{amssymb}

\begin{document}

\begin{frontmatter}




\title{Simulation of the AGILE Gamma-Ray Imaging
Detector Performance: Part I}

\author[label1,label4]{Francesco Longo}

\address[label1]{Universit\`a degli Studi di Ferrara and INFN Sezione di
Ferrara (Italy)}

\author[label2,label4]{Veronica Cocco}
 \address[label2]{Universit\`a degli Studi ``Tor Vergata'' and INFN
Sezione di Roma II, (Italy)}

\author[label3,label2,label4]{Marco Tavani}

\address[label3]{Istituto di Fisica Cosmica, CNR, Milano (Italy) }
\address[label4]{Consorzio Interuniversitario Fisica Spaziale, Torino (Italy)}

\begin{abstract}
We present in this paper the results of a comprehensive GEANT
simulation of the Gamma-Ray Imaging Detector (GRID) being
developed for the AGILE space astrophysics mission. The GRID is
designed to be sensitive in the $\sim 30 $~MeV--50~GeV energy
range, with excellent imaging and timing capabilities,
and with a very large field-of-view ($\sim 3$~sr).
In this paper (Paper I) we present the GRID baseline geometry, a
model for the charged particle and albedo-photon backgrounds
for an equatorial orbit of 550-600 km altitude, and
the main results of the first level (Level-1 Trigger) on-board
data processing.
Our simulations show that the GRID Level-1 data processing is
expected to be capable of decreasing by a factor of $\sim 20$ the
charged particle background (from $\sim 2$~kHz to below 100~Hz),
and by a factor of $\sim
30$ the albedo-photon background. The
gamma-ray photon detection efficiency by the imaging GRID is
simulated to be particularly efficient, varying between 39\% and
17\% depending on photon energies and incident directions.

\end{abstract}


\end{frontmatter}


\section{Introduction}

AGILE is an Italian Space Agency (ASI) Small Scientific Mission
dedicated to high-energy astrophysics \cite{tavaniNIM,tavani-3}.
The AGILE instrument is designed to detect and image photons in
the $\sim 30$~MeV--50 GeV and 10--40 keV energy bands, with
excellent spatial resolution and timing capability, and a
very large field of view covering $\sim$ 1/5 of the
entire sky at energies above 30 MeV \cite{agile-2}. Primary
scientific goals include the study of AGNs, gamma-ray bursts,
Galactic sources, unidentified gamma-ray sources, diffuse
Galactic and extragalactic gamma-ray emission, high-precision
timing studies and Quantum Gravity testing \cite{agile-2}.

The AGILE Gamma-Ray Imaging Detector (GRID) consists of a
Silicon-Tungsten Tracker, a Cesium Iodide Mini-Calorimeter (MC),
an Anticoincidence system (AC) made of segmented plastic
scintillators, and a Data Handling system (DH). The GRID is
sensitive in the energy range $\sim 30$~MeV--50 GeV, and is
designed to achieve an optimal angular resolution (source
location accuracy $\sim$  5'--20' for intense sources), a
very large field-of-view ($\sim 3 $~sr), and a
sensitivity comparable to that of EGRET for on-axis (and
substantially better for off-axis) point sources.
The optimal performance of the AGILE GRID
requires an efficient on-board selection of events aimed 
at maximizing the background rejection and the transmission 
of cosmic photon events. This task is carried out by the Data 
Handling (DH) System.

In the current (Paper I)  and companion \cite{art2} papers we
present the main results concerning the  simulation of GRID
events induced by the particle/albedo-photon background and by the
cosmic gamma-rays. We study in detail different levels of the
GRID on-board trigger and data processing. For the Montecarlo
simulations we used the numerical code GEANT (version 3.21)
\cite{geant} and related subroutines. Our numerical results
 are in  very good
agreement  with CERN beam-test results obtained by our group with
electrons and photons in the 0.1--1~GeV energy range
\cite{barbNIM,tesi-giulio,test-beam}.

\section{\label{par:bkg}Cosmic gamma-ray signal and
expected particle and albedo-photon background}

\subsection{Cosmic gamma-ray flux}

The cosmic photon flux on the AGILE-GRID detector will be dominated by
the diffuse gamma-ray emission, typically larger near the Galactic
plane by  a factor of 10 with respect to the
high-Galactic-latitude. This latter emission has a flux  of $\Phi
\simeq 10^{-5} \, \rm photons/cm^2/s/sr$ for photons above 100
MeV. Expected gamma-ray event rates for the AGILE-GRID depend on
the pointed fields and can be given in the ranges (dominated by
the diffuse gamma-ray emission): 0.02 - 0.1 photons/s for energies 
larger than 100 MeV and 0.1 - 1 photons/s for energies between 10 
and 100 MeV.
The contribution of point-like gamma-ray sources adds to
these rates without changing their order of magnitude.

In the simulations we considered 4 classes of cosmic gamma-ray photons, 
divided in low-energy photons (25--50~MeV) and high-energy photons 
(400--1000~MeV); we used a power-law 
energy spectrum of index -2, and run simulations for two extreme cases: 
an input geometry of incidence angle between 0 and 10 degrees off-axis, and
an input geometry of incidence angle between 50 and 60 degrees off-axis.

\subsection{Charged particle background}

An equatorial orbit near 550 km of altitude for the AGILE mission
will provide a relatively low-background environment.
The charged particle background for this orbit is  known to be 
relatively stable, with an increase by a factor 10-100 near the 
South Atlantic Anomaly (typically influencing about 10\% of the 
orbital duration).
Taking into account data from SAS-2 \cite{fichtel,thompson} and 
Beppo-SAX \cite{sax} missions,
out of the South Atlantic Anomaly we expect an average rate of 
charged particle background above $\sim 1$ MeV of $\sim 0.3 $~particles 
cm$^{-2}$ s$^{-1}$ (mostly electrons/positrons with a
$\sim$10\% contribution of protons).
 
\begin{figure}[!hb]
    \centering
    \includegraphics[width=0.8\textwidth,bb= 90 120 500 435, clip=]{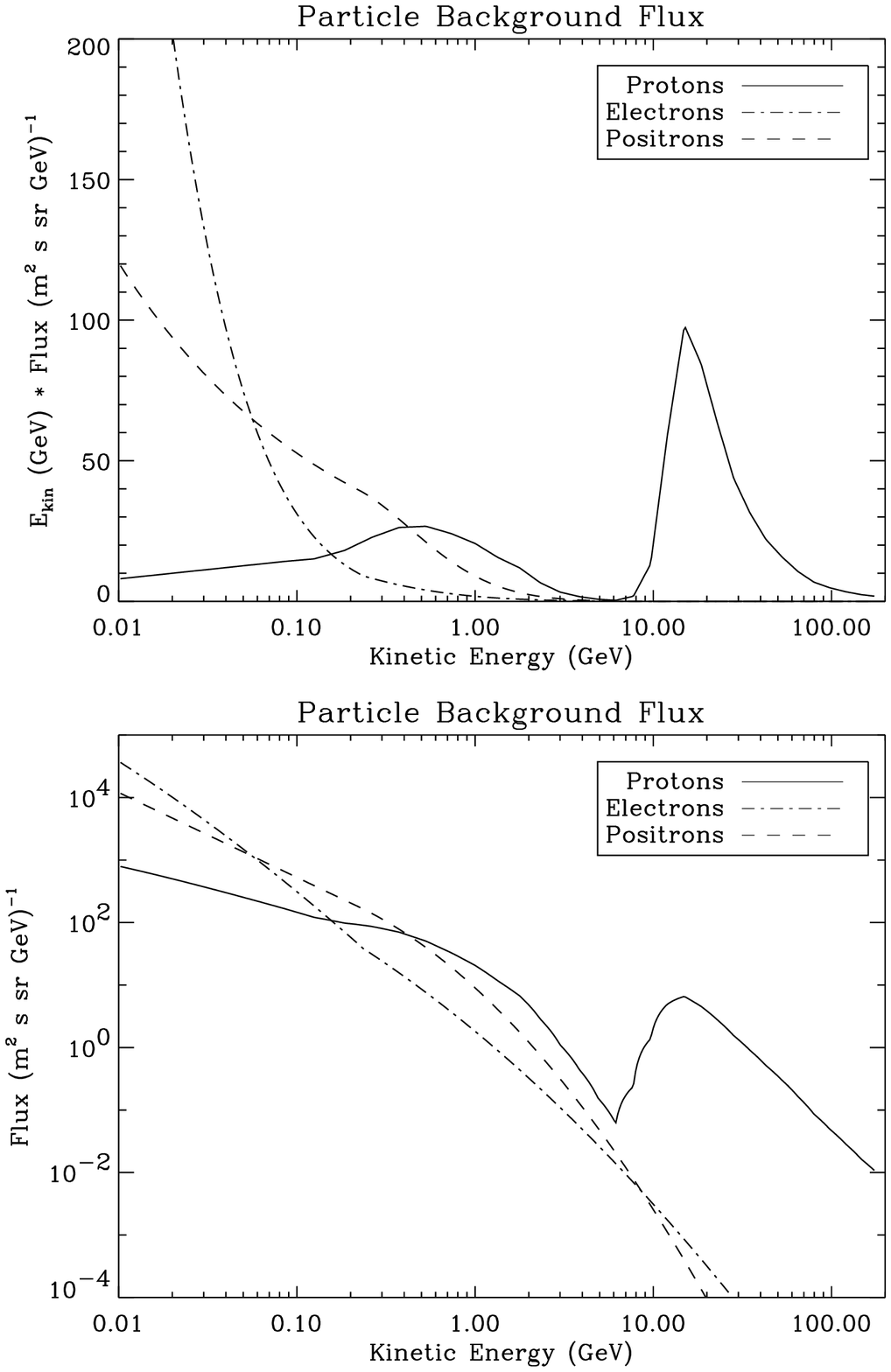}
    \caption{}
    \label{fig:spectra}
\vspace{0.5truecm}
\end{figure}

In our simulations, based on spectral data from AMS 
\cite{battiston,AMS1,AMS2} and MARYA \cite{MARYA}, we
considered the following background components: electrons (ELE),
positrons (POS), primary high-energy protons (HE PROT) and secondary 
low-energy protons (LE PROT).
We made the following assumptions: 1) we  considered an 
``average equatorial position'' and we 
assumed the AGILE z-axis pointed towards the sky for an unocculted
GRID field of view; 2) we considered a spectrum averaged over the 
magnetic latitudes
$|\lambda_{mag}|=0-11^{\circ}$ for secondary protons, and
$|\lambda_{mag}|=0-17^{\circ}$ for electrons and positrons (values compatible
with the AGILE equatorial orbit), regardless of longitude dependence;
3) we considered an isotropic distribution on a virtual sphere containing 
the satellite for particles below the geomagnetic cutoff (albedo leptons 
and secondary protons), and an isotropic distribution emitted from
the upper hemisphere for the primary protons above the cutoff.
The adopted particle spectra, which are fits of the AMS and MARYA
data (calculated in \cite{lipari-longo}) are shown in
Fig.~\ref{fig:spectra}. Their fluxes are normalized to the values
indicated in Tab.~\ref{tab:flux} and the total incidence rate on
a surface is assumed to be  $\sim$~0.3 particles~$\rm cm^{-2} \,
s^{-1}$ for particle kinetic energies above $\sim 10$~MeV.

\subsection{Albedo gamma-ray background}

The interaction of the charged cosmic-rays with the upper atmosphere
induces a relatively strong and non-isotropic gamma-ray background. It peaks 
near the Earth horizon (corresponding to a zenith angle 
$\theta = 112^{\circ}$ for an orbit of 550~km altitude), and has
a characteristic East-West asymmetry (by a factor of $\sim 4$ in
intensity near 40 MeV) \cite{albphot}.

In our simulations, we used a simplified but representative model of the
albedo-photon emission, and we considered two geometries:
(1) albedo photons reaching the GRID "from below", a situation valid for
the unocculted portions of the AGILE orbit (ALB-1 PHOT);
(2) Earth covering half of the AGILE FOV, with a large fraction of
  albedo photons reaching the GRID "from one lateral side" (ALB-2 PHOT).
Based on balloon flight data
\cite{albphot4,ling,matteson,albphot2,albphot3} and SAS-2 data
\cite{albphot}, we have assumed the following simplified spectrum
representing the average of the emission over the whole solid
angle subtended by the Earth surface ($\sim 4$ steradians) at the
height of 550~km:
\begin{eqnarray}
    \frac{dN}{dS\;dt\;dE\;d\Omega}=\left\{
            \begin{array}{ll}
           0.08\;E^{-1.4}\; \rm cm^{-2}s^{-1}sr^{-1}MeV^{-1} & \mbox{for $\rm 1\:
           MeV<E<10\: MeV$} \\
            0.5\;E^{-2.2}\;\; \rm cm^{-2}s^{-1}sr^{-1}MeV^{-1}  & \mbox{for $\rm 10\:
           MeV<E<100\: GeV$}
        \end{array}
        \right.
        \nonumber
\end{eqnarray}
with a total flux ${\mathcal F}= 0.15 \: \rm cm^{-2}s^{-1}sr^{-1}$ in the
energy range 1 MeV - 100 GeV.

\subsection{Background rejection}

The ratio of charged particle events (penetrating the AC) to cosmic
gamma-ray photon events is typically of order $10^3 - 10^4$ for
photons of energy 20-100 MeV, and $10^4-10^5$ for photons above
100 MeV. Albedo photons are larger than cosmic gamma-ray photons
by factors between 10 and 100 (depending on the pointing geometry
and energy range).

Clearly, is necessary to `filter' the GRID events.
This task is carried out by  a hardware-implemented
fast logic (``Level-1 trigger"), and by a set of asynchronous
software algorithms and CPU processing ("Level-2 processing").
The adopted requirements for the DH processing
depend on the  downlink telemetry rate, assumed
to be 512 kbit/s during the 10 minute duration of satellite
visibility from the ASI ground station in Malindi (Kenya).
A main aim of  the AGILE DH system is to provide an on-board
filtering of events  reducing the background rate
to an acceptable value  within a factor of 10--100 of the cosmic
gamma-ray photon rate. A final filtering of the particle and
albedo-photon background will be carried out on the ground by a
dedicated software.
The charged particle and the albedo-photon background models described in
the previous sections were used in Montecarlo simulations in order to optimize
the DH processing.

\section{The AGILE Model and GRID detection assumptions}

\subsection{The AGILE Model}
\begin{figure}[h]
\centering
\includegraphics[width=0.75\textwidth,bb=60 108 508 507,clip=]{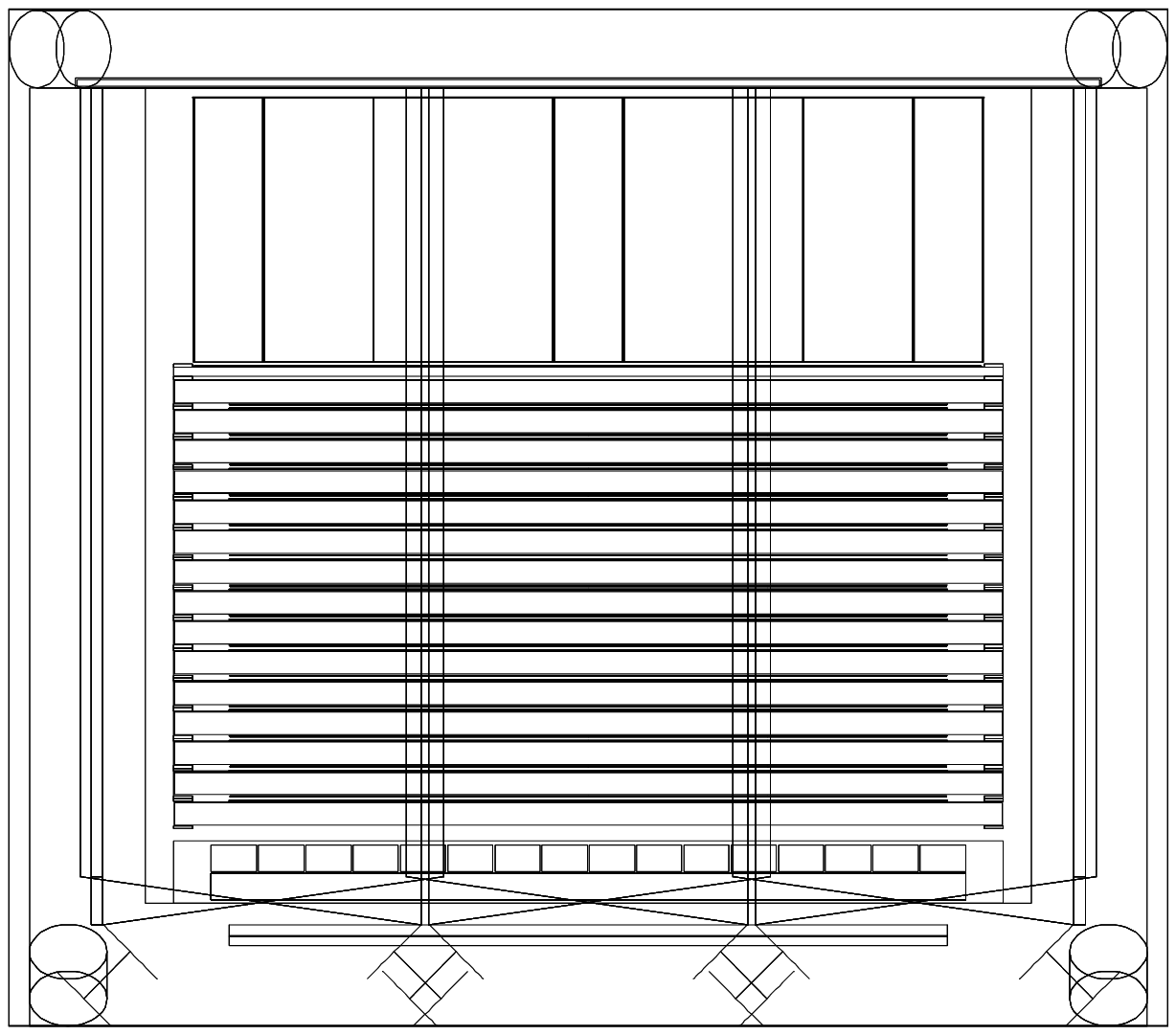}
\caption{\label{fig:figure2}{\small}}
\vspace{0.5 truecm}
\end{figure}

\begin{figure}[hb!]
\centering
\includegraphics[width=0.8\textwidth,bb=62 54 515 514,clip=]{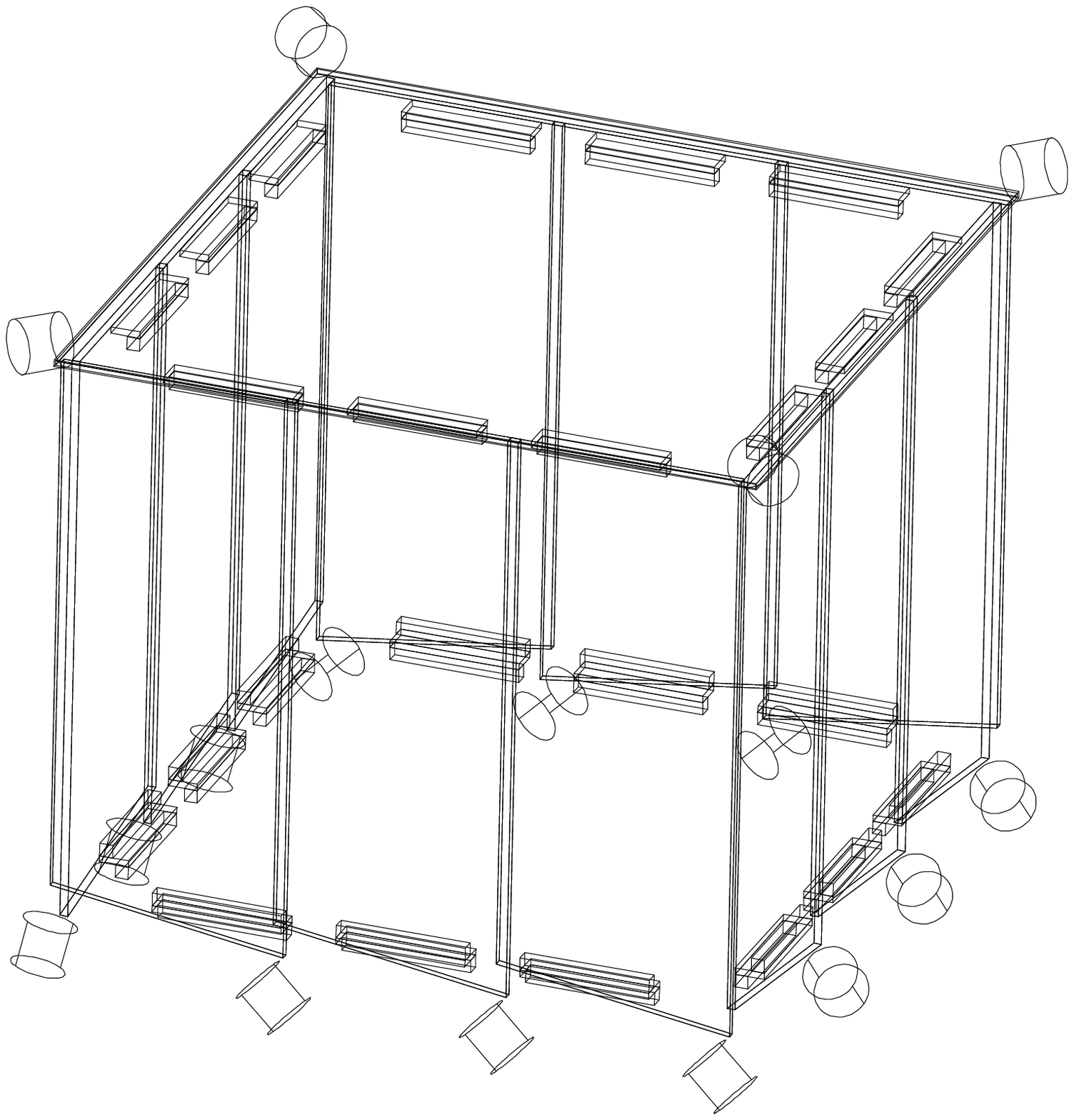}
\caption{\label{fig:acd}{\small}}
\vspace{0.5truecm}
\end{figure}

\begin{figure}[ht!]
\centering
\includegraphics[width=0.8\linewidth,bb=110 240 450 465,clip=]{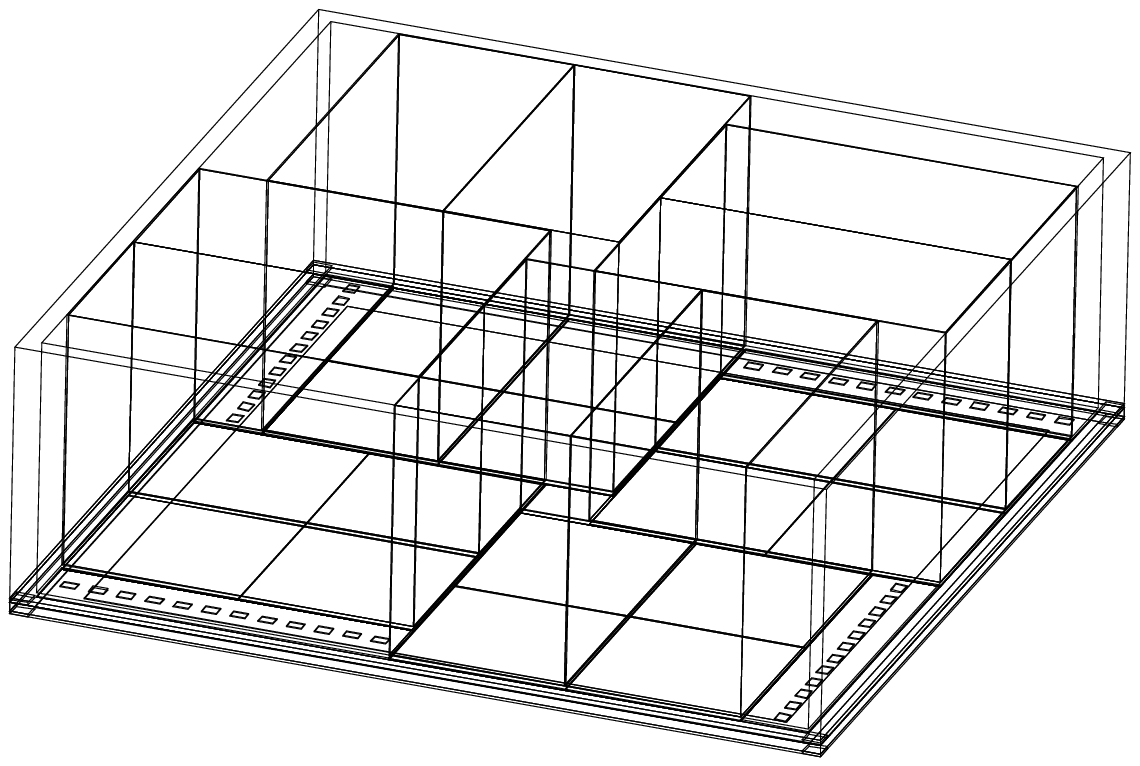}
\caption{\label{fig:sup4}{\small }}
\vspace{0.5truecm}
\end{figure}

A correct mathematical and physical model  of the AGILE payload and spacecraft
with all their relevant components is crucial to evaluate the GRID behaviour
and performance.
A general description of the AGILE instrument can be found in
\cite{tavaniNIM,agile-2}.
Simulation programs consider the following AGILE components
(see Fig. \ref{fig:figure2}): the spacecraft (bus) MITA and the AGILE payload,
 consisting of the CsI Mini-Calorimeter, the Silicon-Tungsten Tracker, the
Anticoincidence system, the X-ray detector (Super-AGILE), the thermal shield,
the Mechanical structure and the lateral electronics boards.
The  AGILE components were modeled as follows:
\begin{itemize}
\item[{\bf 1)}] {\bf MITA spacecraft}\\
The spacecraft (bus) MITA is represented by a box, made of Carbon-fibers,
containing 10 equidistant Aluminum layers inserted in thin (100 $\rm \mu m$)
Carbon-fiber structures; every Aluminum layer is 1.1  cm  thick for a total
weight of 128 kg.
\item[{\bf 2)}] {\bf Mini-Calorimeter}\\
The Mini-Calorimeter is modeled by two pla\-nes,
each containing 16 CsI bars, oriented as the y-axis
in the upper plane and as the x-axis in the lower plane.
The CsI bars are 1.4 cm thick and have a
pitch of 2.4 cm. They are inserted in a Carbon-fiber structure 0.1
cm apart one each other. The distance between the X layer and the Y layer is
assumed to be 0.1  cm. The Mini-Calorimeter is placed directly above the
bus MITA, at a 0.5  cm distance from the Silicon Tracker.
\item[{\bf 3)}] {\bf Silicon Tracker} \\
The Silicon Tracker is made of 14 detection planes,
with a distance between consecutive planes of 1.6 cm and a distance between
the X and Y layer in each plane of 0.16 cm; the first 12 planes contain
also a Tungsten layer each ($245\; \rm \mu m$ thick).
Each Silicon layer is composed of 16 tiles of $9.5\times 9.5\;
\rm cm^{2}$ each, with a pitch of $121 \rm  \mu m$ (each tile
contains 768 strips) and a thickness of $410\; \rm  \mu m$. In
our model there are also the honeycomb support, the Aluminum
inserts connecting the Tracker planes with the main structure of
the satellite, and the Front-End electronics chips (3 TAA1 for
each ladder). Every Tracker plane is composed by a Carbon-fiber
structure ($500 \; \rm  \mu m$ thick) supporting an Aluminum
honeycomb. Under the Carbon-fiber are glued the tungsten layer
(for the first 12 planes), one layer of Kapton ($100 \; \rm  \mu
m$ thick) with a Copper coating ($15 \; \rm  \mu m$ thick) and
the X Silicon layer. Under the X layer there are the Y layer and
another Kapton-Copper layer, which in the reality are glued to
the next plane structure. On the same structure are glued also
the ceramic hybrids which sustain the TAA1 Silicon chips.
The global structure is composed by 14 X and Y detectors layers
and by 15 structure supports; the total Tracker height is 24 cm.
\item[{\bf 4)}] {\bf Anticoincidence system}\\
 The AC system is made of a top panel of
 plastic scintillator
($54\times 54 \times 0.5\; \rm cm^{3}$) and 3 overlapping
panels for each side of the AGILE Tracker (0.6 cm thick, 18.1 cm
large and  44.4 cm high). The AC plastic panels are inserted in a
Carbon-fiber structure ($500 \; \rm  \mu m$ thick) and are
supported by some Aluminum inserts; the trapezoidal form of the
lateral panels is described in a proper way. We have inserted
also a schematic description of the photomultipliers and their casings.
  The AC system surrounds the Mini-Calorimeter, the Silicon Tracker and
Super-AGILE, and is positioned above the bus MITA.
Fig.~\ref{fig:acd} shows the adopted model.
\item[{\bf 5)}] {\bf X-ray detector (Super-AGILE)}\\
The X-ray detector is made substantially by these elements:
a Silicon detection layer, a collimation system, a gold mask.
A Tungsten ring is positioned below the Si detection plane in order 
to reduce the diffuse photon background.
The detection layer is divided in 16 tiles with the same characteristics 
of the Tracker tiles, described in a realistic way, with the Silicon 
microstrips and the FrontEnd electronics.
We have modeled also the honeycomb and Carbon-fiber support structure; 
the detectors and the support of the coded mask are glued on this structure.
 The Silicon detector is placed over the Tracker, while the distance between 
the Silicon detector and the gold mask ($90 \; \rm  \mu m$ thick) is fixed 
to be 14 cm.
Also the collimation system, made of an ultra-light Carbon-fiber structure
($500 \; \rm  \mu m$ thick), is described in great detail (Fig.~\ref{fig:sup4}).
Its internal panels are coated with a $75 \; \rm  \mu m$ thick gold layer, 
in order to reduce the contribution from the diffuse cosmic X-ray background. 
The distance between the gold mask and the top AC panel is 0.5 cm.
\item[{\bf 6)}] {\bf Thermal shield}\\
The thermal shield is modeled as an external Teflon layer  (1 mm thick) plus an
internal Aluminum layer (10 $\rm \mu m $ thick).
It covers the Mini-Calorimeter, the Tracker, Super-AGILE and the AC
system.
\item[{\bf 6)}] {\bf Mechanical Structure}\\
The description of the payload is completed by a preliminary model for
the general support structure: four Aluminum legs connected by
other Aluminum bars. \\
There is also a preliminary description of the Front End Electronics (FEE)
and of the trigger vertical boards,
one for each lateral side of the payload, and one for the DH box, under the
Mini-Calorimeter. They are described as some layers of Silicon, ceramics
and Copper.
\end{itemize}

\subsection{Silicon Tracker capacitive coupling,
floating strip readout, and Gaussian noise}

The electronic system for the AGILE Silicon Tracker is based on the
``floating strip'' readout.
This implies that only one out of two contiguous Si-microstrips (each of 121
$\rm \mu  m$ size) is read by dedicated electronic devices (TAA1 chips).
However, the capacitive coupling between contiguous strips allows not to loose 
any information.
In the Montecarlo code the capacitive coupling is simulated calculating
the total energy release in every readout strip by the weighted
sum as obtained in Ref. \cite{tesi-giulio}.
For the $k$-th readout strip, the total energy release taking into account 
neighbour strips is
$
E_s(k) = E_{i}+0.38 \cdot (E_{i-1}+ E_{i+1})+0.115 \cdot (E_{i-2}+ E_{i+2})+  
          0.095 \cdot (E_{i-3}+ E_{i+3})+ 0.045 \cdot (E_{i-4}+ E_{i+4})+ 
          0.035 \cdot (E_{i-5}+ E_{i+5}) $,
where $E_{i}$ is the energy release in the $i$-th strip that can be either 
readout or floating, and $k=1+(i-1)*2$. 
This scheme for the capacitive coupling simulation is in agreement with 
test-beam results obtained from our group for a variety of
incidence angles \cite{barbNIM}.

Instrumental noise was simulated as a Gaussian distribution
with $\sigma=5\:$~keV.
We found that for a detection threshold of 1/4 MIP$\simeq 27\:$keV it has no
macroscopic effects on the trigger rates.
All the simulation results presented in this document have been
obtained using the geometry model described above, taking into account
the capacitive coupling and the floating strip readout, but without 
considering the noise that has negligible effects on the results.

\section{Level-1 trigger}

The Level-1 trigger for the Agile GRID is a hardware-implemented fast logic
required to be very rapid ($\sim$ 2 $ \rm \mu s$), conceptually simple,
easily implementable by a dedicated hardware, and using parameters 
reconfigurable by Telecommands.
With Montecarlo simulations we tested different trigger configurations, using 
the event classes (cosmic gamma-ray photons, background particles and Earth
albedo photons) described in section~\ref{par:bkg}.
Here we present the main results.

Our best strategy for the Level-1 logic
 is based on the combined use of
signals from AC panels and of the ratio R,
defined as the ratio between the total number of fired TAA1 and the
total number of fired X and Y views:
$R=(\mbox {total n. of fired TAA1})/(\mbox {total n. of fired X
     and Y views})$.
In what follows,
we call "$R$--trigger"
 the logic (implemented by a dedicated electronic component)
regulating the use of both AC panels and $R$ threshold values.

\subsection{\label{par:trg1def}Level-1 trigger configurations}

Here is a short description of the different Level-1 trigger
configurations that we studied:
\begin{itemize}
    \item{\bf PLA} = events which give hits in at least
    3 out of 4 consecutive planes (X OR Y view)
    \item{\bf TOP} = PLA events which pass the TOP AC veto
    \item{\bf LAT} = TOP events with signals in 0 or 1 lateral AC panels,
    and TOP events which give signals in 2 consecutive AC panels or
    in 2 AC panels on the same side
    \item{\bf LSI} = LAT events with signals in 0 lateral AC panels,
    and LAT events with signals in 1 or 2 AC panels but with NO signal
    in the last silicon plane
    \item{\bf R11G} = LAT events with signals in 0 lateral AC panels,
    and LAT events with signals in 1 or 2 AC panels and $R>1.1$
\end{itemize}
PLA, TOP and LAT are consecutive steps, while LSI and R11G are
the two final alternative trigger configurations.
For comparison, in Tables \ref{tab:trig1fondo}, \ref{tab:trig1albedo} 
and \ref{tab:trig1fot} we reported also the number of events
characterized by primary particles or photons reaching
 the Tracker volume (TRA), a box of
$38.06\times 38.06\times 21.078$ $\rm cm^{3}$ which includes the
Tracker planes from the top sheet of the first tungsten layer to
the bottom sheet of the last Silicon-y plane.

In the case of gamma-ray photons it is important to observe that:
\begin{itemize}
    \item [1)] gamma-photons giving a signal in the Tracker may
    convert in the Tracker volume, or in the Super-AGILE volume
    (mask or collimators). They can also convert elsewhere (mechanical
    structure, Mini-Calorimeter or bus MITA), or pass through the
    instrument without converting. We define as ``good photons'', those for
    which there is  high probability to reconstruct the incident
    direction. By definition they are ``Tracker-converted photons'';
    \item[2)]because of the presence of the Super-AGILE structure,
     some gamma-photons potentially able to enter into the
     Si-Tracker fiducial volume, because of their incidence angles
     and energies,
     can convert in the Super-AGILE structure.
      Their secondaries (electron
    and positron) may never arrive into the Tracker volume. On the
    other hand, it can happen that some  photons, that geometrically
    would not enter into the Tracker
    volume, convert in the Super-AGILE structure,
     and their secondaries give a signal in
    the Tracker then inducing a GRID background.
\end{itemize}

It is then useful to indicate with the suffix ``TC''  the photons
converted in the Tracker volume, and  with the suffix ``SC'' the
photons converted in the Super-AGILE structure.
In the case of gamma-ray photons we also distinguish: a) photons that theoretically, 
from a geometrical point of view, might enter into the Tracker volume (TRA-TH),
b) primary photons actually entered in the Tracker volume (PTRA), c) primary photons 
converted in the Tracker volume (PTCON).

\subsection{Results and discussion}
\begin{figure}[!h]
\begin{center}
\includegraphics[width=\textwidth]{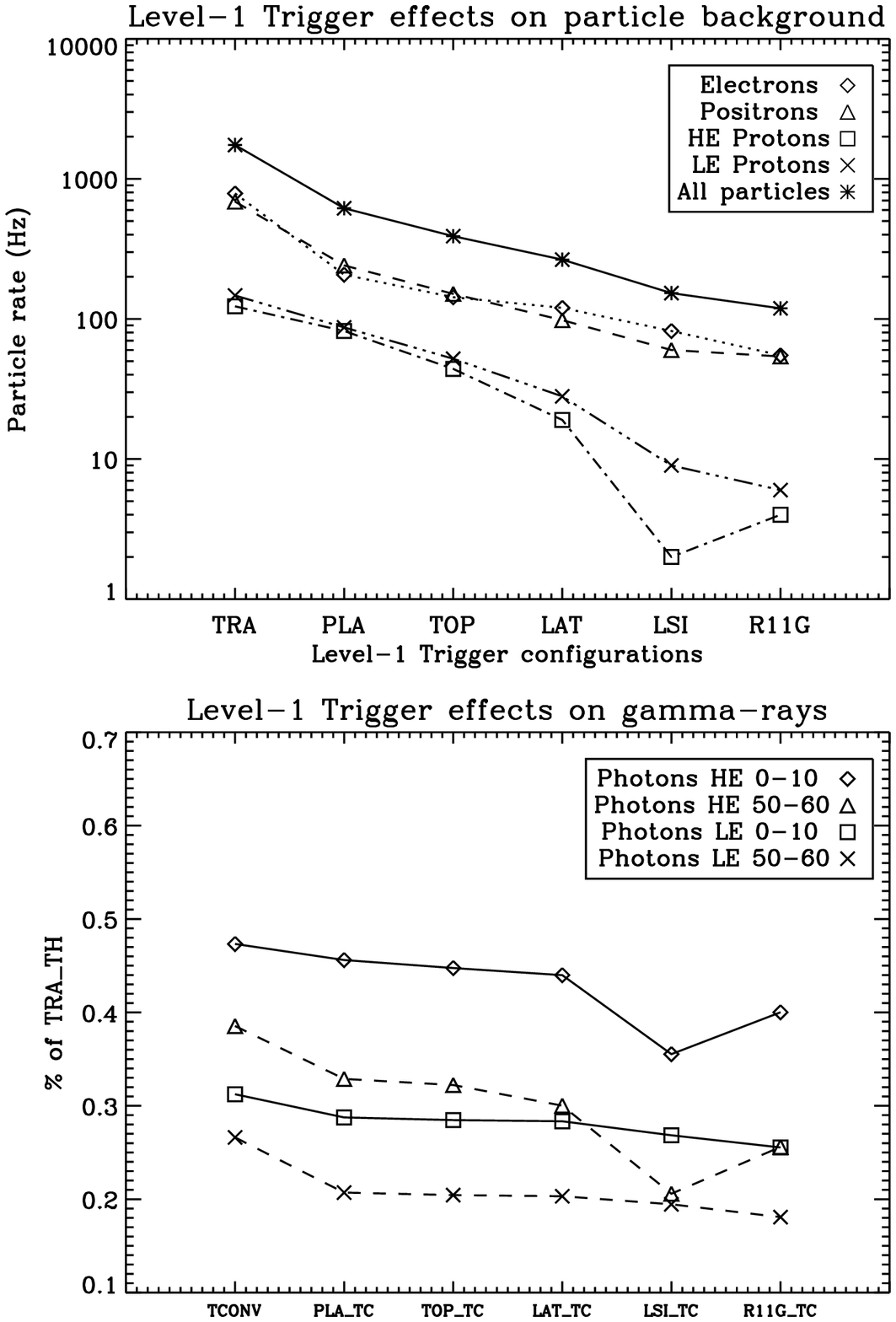}
\caption{\label{fig:trg1}{\small}}
\end{center}
\vspace{0.5truecm}
\end{figure}

Tabs.~\ref{tab:trig1fondo}, \ref{tab:trig1albedo} and \ref{tab:trig1fot}
and  Fig.~\ref{fig:trg1}
summarize our results  in terms
of alternative or progressive event selections.
The TOP AC veto is useful to reject downgoing charged particles;
the LAT AC veto discriminates events giving a signal either
entering than exiting the detector (charged particles) from
events which give signals only when they escape from the Tracker
(photons into charged pairs).
The R-trigger is a method optimized to reject non-interacting charged
particles, based on a simplified idealization:
 when
a non-interacting particle enters in the Tracker, only one track is expected;
if the detector is crossed by a photon  creating an electron-positron
pair, two tracks are expected.
Simulation results show that the R-trigger works very well with
protons, and gives acceptable results also for electrons and
positrons. In general, the
method succeeds in discriminating photons from charged background
particles very efficiently, sometimes better than the AC system.
Clearly, the use of the $R$-trigger does not reject protons
generating many secondaries or e$^{+}$/e$^{-}$ producing
secondaries and electromagnetic showers. A trigger configuration
based only on the AC system does not reject low energy protons
stopping inside the Tracker.
The two methods are based on different ``philosophies'' and they both
give good results in different situations.

From our simulations it appears that the best choice
is a combination of the two strategies and that
the most efficient Level-1 trigger
configuration is R11G, rejecting $\sim 93 \%$ of
background particles
without affecting significantly the gamma-ray detection rate.

\section{Level-1.5 Data Processing}

The additional ``Level-1.5''  data processing uses the Tracker
discretization in terms of TAA1 chips ($3.1\times 3.1$ $\rm
cm^{2}$ square elements), together with the complete AC  and MC
information.
We investigated the efficiency of alternative Level-1.5 Trigger
options in rejecting background charged particles and here we
present the main results.

\subsection{Level 1.5 trigger configurations}

We considered  the following options:

\begin{itemize}
    \item{\bf DIS =}The DIS option is a simplified track
    reconstruction and is based on computing the distance $D$ of the
    fired TAA1s  from the fired AC lateral panel.
    The basic
    idea is that a charged particle is expected to have  increasing
    $D$ values as a function of increasing plane, while a gamma-ray photon is
    expected to have a decreasing $D$ function (in the restricting
    hypothesis that they both come downward and in the absence of
    strong ``hard'' scattering of photon-created pairs). The parameter DIS is
    defined as: $DIS=D_{fp}-D_{lp}$ where $D_{fp}$ is the distance of the closest
    fired TAA1 to the fired AC lateral panel in the first
        plane, while $D_{lp}$ is the distance of the closest
    fired TAA1 to the fired AC lateral panel in the last
        plane.
    We require $DIS\geq 0$ for good events. This option is applied 
    only if there are fired AC lateral panels.
    \item{\bf RUD10 (or RUD12) =} The fired Tracker views are divided in two
    groups (UP and DOWN) and the ratio R is computed separately for
    the two groups ($R^{up}$ and $R^{down}$). The ratio
    $RUD=R^{up}/R^{down}$ can help in discriminating good gamma-photons,
    that come downward, from secondary photons generated by charged
    particles, that come from the bottom toward the top of the detector.
    With the option RUD10 are rejected all the events with $RUD>1.0$,
    while with the option RUD12 are rejected all the events with
    $RUD>1.2$.
    \item{\bf VFORM =} This option, as RUD10 and RUD12, can help in
    discriminating primary photons going downward from secondary
    photons coming upward. It computes the separation $S$ between fired
    TAA1 on the same view, divides the fired planes in two groups (UP
    and DOWN) and compares the mean separation of the UP
    planes with the mean separation of the DOWN planes. The
    parameter $V$ is defined as $V=S_{down}-S_{up}$ and events with
    $V<0$ are rejected.
    \item{\bf CGAP =} This option rejects events with a gap of
    more than 2 planes (4 views) between the MC and the last fired
    Tracker view. It is applied only if there is
    a signal from the MC
    and it is supposed to reject events due to $e^{+}$ or $e^{-}$ which
    stop in the MC and generate a secondary photon which enters the
    Tracker upward and creates a couple only after passing 4 or more
    views.
    \item{\bf TGAP =} This option rejects events with a gap of
    more than 2 planes (4 views) in the Tracker. The parameter T\_GAP
    is defined, for a single event, as the maximum number of no-fired
    views between two non consecutive fired views.
\end{itemize}

\subsection{Results and Discussion on the Level-1.5 Trigger Processing}

\begin{figure}[!h]
    \centering
    \includegraphics[width=\textwidth]{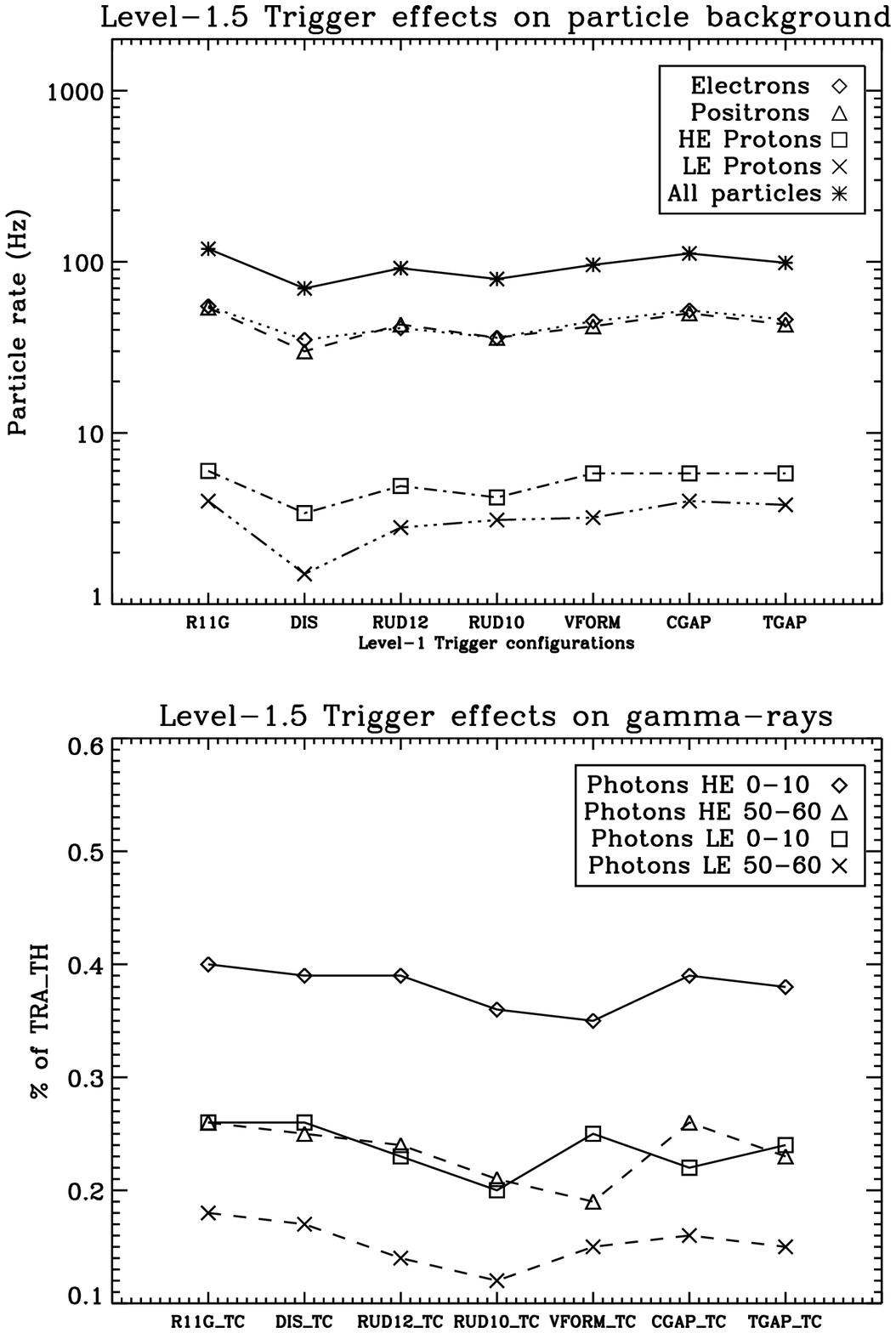}
    \caption{{\small }}
    \label{fig:trig15}
\vspace{0.5 truecm}
\end{figure}

Tables~\ref{tab:trig15} and~\ref{tab:trig15-2}
 and Fig.~\ref{fig:trig15} show the simulation results.
The simple DIS algorithm is the most efficient in rejecting
background particles without losing too many good photons. This
algorithm is applied only if there is at least one fired AC
lateral panel; most of background charged particles verify this
condition, while good photons often do not hit any lateral AC panel.

Options RUD10, RUD12 and VFORM are not very efficient because the
distributions of the RUD and V parameters are not significantly
different for background particles and for good photons. The
C\_GAP option is not efficient because the majority of background
electrons, positrons and albedo protons do not give a signal in
the Mini-Calorimeter. There are no significant differences between
the background particle T\_GAP distributions and the photon
T\_GAP distribution, probably because this parameter is
influenced by secondary particles.

\section{Discussion and Conclusions}

By using a complete and detailed model of the AGILE instrument, we
simulated with GEANT the performance of the on-board Level-1
trigger logic.
 We obtain satisfactory results from our optimization study of the 
Level-1~and~1.5 trigger logic for the cosmic gamma-ray signal (with photon
energies between $\sim 30$~MeV and 50~GeV) and for the rejection
of the particle and albedo-photon backgrounds in a 550 km
equatorial orbit.
The optimal Level-1 (and therefore
necessarily  simple) event selection algorithms are
 the R11G and DIS procedures that can be
easily hardware-implemented and are very fast ($\leq 20\mu$s).
Table~\ref{tab:trig15s} summarizes the main results of our paper:
the total background rates expected for two
different Earth occultation geometries. 
At the Level-1$+$1.5 the
background component completely dominates the event rate over the
cosmic gamma-rays by a factor between 10 and 100, and
the total background
event rate is near 100~$\rm s^{-1}$, a value which can be sustained 
by the on-board AGILE Data Handling \cite{tavani-3}.
Therefore the Level-1
logic presented here is adequate for the AGILE GRID.

We also note that the albedo photons contribute a significant
fraction (between $\sim 30$\% and more than $\sim 50$\% depending
on the Earth's position with respect with the GRID) of the total
background rate. Since these photons are dominated by  10--100~MeV
gamma-rays, Level-1 techniques are not adequate in substantially
reducing their event rate, and a special Level-2 data processing
is required,which will be the subject of a forthcoming paper.

Our results are of general interest for space detectors similar to
the AGILE GRID. Two conclusions are of particular interest: (1)
the fact that low-energy leptons (magnetospheric trapped
electrons and positrons) dominate the particle background requires
special attention and a proper event selection logic; (2) Earth
albedo-photons constitute a substantial fraction of the residual
events passing the Level-1 trigger stage.

\section{Acknowledgments}

Results presented in this paper are based on joint work with
members of the AGILE Team. In particular, we warmly thank
G.~Barbiellini, P.~Picozza and A.~Morselli for special support,
M.~Prest, E.~Vallazza and G.~Fedel for exchange of information on
the Silicon Tracker performance and experimental data, and the
AGILE Simulation and Theory Group for many discussions. We also
thank A.~Pellizzoni and P.~Lipari for collaboration on several
background source function algorithms used in our simulations.

The current  work was carried out at the  University of Rome "Tor
Vergata", University of Ferrara and CNR and INFN laboratories,
under the auspices and partial support of the Agenzia Spaziale
Italiana.

\vspace{2 cm}
{\bf FIGURE CAPTIONS:}\\
\\
Fig.1: Charged particle background for an equatorial
    orbit of 550 km  (results from ref.~\cite{lipari-longo})
     that we assumed in our calculations.
    Electron, positron and proton spectra were obtained by combined AMS-Shuttle
    Flight data and MARYA data. All particle components have an
    approximately isotropic distribution of incidence angles on
    the GRID, except the primary high-energy protons above the geomagnetic
    cutoff of $\sim 6$~GeV.\\
Fig.2:  AGILE payload: from the bottom the different
volumes are:  the DH box, the 2 pla\-nes of the  Mini-Calorimeter, the 14
layers of the Silicon Tracker, Super-AGILE, the top AC panel.
Also the lateral AC panels with the photomultipliers and
the thermal shield are visible. The overall dimensions are: length of 62.55 
cm, width of 62.55 cm and  height of 54.02 cm. The GRID active volume 
dimensions are:
$38.06\times 38.06\times 21.078$ $\rm cm^{3}$  including the
Tracker planes from the top sheet of the first Tungsten layer to
the bottom sheet of the last Silicon-y plane.\\
Fig.3: Anticoincidence system. The overall dimensions are:  lenght of 62.25 cm; width of 62.25 cm; height of 53.87 cm. \\
Fig.4: Super-AGILE. Overall dimensions: lenght of 44 cm; width of 44 cm; height of 14.77 cm.\\
Fig.5: Particle background rates for different Level-1 trigger 
configurations and Level-1 trigger efficiency for different photon classes.
Event selection based on TRA, PLA, TOP and LAT are sequentially
applied to all events.
Further selections based on LSI or R11G are instead mutually exclusive.
 The suffix ``TC'' denotes ``Tracker-converted''.\\
Fig.6: These plots show the efficiency of
the Level-1.5 Trigger alternative options in rejecting particle
background and in detecting gamma-photons (DISRUD means (DIS+RUD\_10)
and DISCG means (DIS+C\_GAP)).

\renewcommand{\baselinestretch}{1}
\vspace{2 cm}
\begin{table}[!ht]
\begin{center}
\caption{\label{tab:flux}{\small {\bf Integrated Particle Flux
Rates}}} \vskip .08in \small
\begin{tabular}{ l  r r  c c }
\hline
Particles & $E_{min}$ & $E_{max}$ & ${\mathcal F}$ ($ \rm cm^{-2}s^{-1}sr^{-1}$) &
$\pi$ ${\mathcal F}$ ($ \rm cm^{-2}s^{-1}$)\\
\hline
Electrons   &   10 MeV  &   10 GeV  &    3.76  $10^{-2}$  &   0.1180 \\
Positrons   &   10 MeV  &   10 GeV  &    2.67 $10^{-2}$ &   0.0840 \\
Albedo Protons &  10 MeV &  6 GeV   &    9.04 $10^{-3}$ &   0.0284 \\
Primary Protons &  6 GeV & 180 GeV  &    9.99 $10^{-3}$ &   0.0314 \\
\hline
\end{tabular}
\end{center}
\end{table}
\vspace{1cm}

\begin{table}[!ht]
\begin{center}
\small \caption{\label{tab:trig1fondo}{\bf
Level-1 Trigger Selection for Background Charged Particles}}
\vskip .08in
\begin{tabular}{l c c c c c}
\hline
        & ELE & POS  & HE PROT & LE PROT & TOTAL  \\
 &($\rm s^{-1}$) & ($\rm s^{-1}$) & ($\rm s^{-1}$) & ($\rm s^{-1}$)& ($\rm s^{-1}$)\\
 & (percent.) & (percent.)  & (percent.)  & (percent.) & (percent.) \\
\hline
TRA           &  787    &  688    &  123   &  147   &  1745 \\
\hline
PLA           &  207    &  241    &  82    &  87    &   617 \\
(\% of TRA)   &   26\%  &  35\%   &  67\%  &  59\%  &   35\% \\
\hline
TOP           &  143    & 151     &  44    &  52    &   390 \\
(\% of TRA)   &  18\%   & 22\%    &  36\%  &  35\%  &   22\% \\
\hline
LAT           &  120    &  98     &  19    &  28    &  265  \\
(\% of TRA)   &  15\%   &  14\%   &  15\%  &  19\%  &   15\% \\
\hline
LSI           &  82     &  60     &   2    &   9    &  153 \\
(\% of TRA)   &  10\%   &   9\%   &  1\%   &   6\%  &   9\% \\
\hline
R11G          &  55     &  54     &   4    &   6    &  119 \\
(\% of TRA)   &  7\%    &  8\%    &  3\%   &   4\%  &    7\% \\
\hline
\end{tabular}
\end{center}
\vspace{0.5 cm}
\end{table}
\begin{table}[!ht]
\begin{center}
\small \caption{\label{tab:trig1albedo}{\bf
Level-1 Trigger Selection for Background Albedo Photons}} \vskip
.08in
\begin{tabular}{l c c }
\hline
        & ALB-1 PHOT    &   ALB-2 PHOT    \\
        &($\rm s^{-1}$) & ($\rm s^{-1}$)  \\
\hline
TRA           &  748    &  1292     \\
\hline
PLA           &   30    &    48    \\
\hline
TOP           &   25    &   45     \\
\hline
LAT           &   24    &   44    \\
\hline
LSI           &   22    &   41     \\
\hline
R11G          &   22    &   39     \\
\hline
\end{tabular}
\end{center}
\vspace{0.5 cm}
\end{table}
\begin{table}[!t]
\begin{center}
\small \caption{\label{tab:trig1fot}{\bf
Level-1 Trigger Selection for Cosmic Gamma-Rays (*)}}
\vskip .08in
\begin{tabular}{l c c c c }
\hline
              & PHOT HE 0-10 & PHOT HE 50-60  & PHOT LE 0-10  & PHOT
              LE 50-60 \\
 &(event no.) & (event no.) & (event no.) & (event no.)\\
\hline
TRA\_TH      &  3144   &  3352   &  3144  &  3352   \\
\hline
TRA           &  3511   &  3744   &  3433  &  3646   \\
PTRA          &  2975   &  3054   &  3009  &  3175   \\
PTCON         &  1488 (47\%)  &  1292 (39\%)  &   982 (31\%)  &   892 (27\%)   \\
\hline
PLA           &  1683   &  1415   &  1171  &   908   \\
PLA\_TC       &  1434 (46\%)  &  1102 (33\%)  &   904 (29\%) &   694 (21\%)  \\
PLA\_SC       &   168   &   128   &    95  &   44    \\
\hline
TOP           &  1608   &  1355   &  1134  &  882   \\
TOP\_TC       &  1407 (45\%)  &  1080 (32\%)   &   895 (28\%)  &  685 (20\%)   \\
TOP\_SC       &   156   &   121   &    91  &   40   \\
\hline
LAT           &  1578   &  1213   &  1129  &  872   \\
LAT\_TC       &  1383 (44\%)  &  1006 (30\%)  &   891 (28\%) &  681 (20\%)   \\
LAT\_SC       &   150   &   102   &    90  &   40   \\
\hline
LSI           &  1243   &   829   &  1071  &  833    \\
LSI\_TC       &  1117 (36\%)  &   690 (20\%)   &   844 (27\%)  &  652 (19\%)   \\
LSI\_SC       &    94   &    77   &    89  &   40   \\
\hline
R11G          &  1469   &  1029   &   998  &  766   \\
R11G\_TC      &  1258 (40\%)  &   858 (26\%)  &   803 (26\%) &  606 (18\%)   \\
R11G\_SC      &   136   &    84   &    72  &   36   \\
\hline
\end{tabular}
\end{center}
\vspace{0.3truecm}
(*) In parenthesis we reported the percentage of selected events respect to the total number
 of photons that theoretically could enter into the Tracker volume (\% of TRA\_TH)
\vspace{0.5truecm}
\end{table}

\begin{table}[!h]
\begin{center}
\caption{\label{tab:trig15}
{\bf Summary of Level-1.5  processing of background events}}

\begin{tabular}{l c c c c c c c}
\hline
  Type     & R11G & DIS & RUD12 & RUD10 & VFORM & CGAP &  TGAP  \\
&($\rm s^{-1}$)&($\rm s^{-1}$)&($\rm s^{-1}$)&($\rm s^{-1}$)&($\rm s^{-1}$)
&($\rm s^{-1}$)&($\rm s^{-1}$) \\
  \hline
ELE  & 55 &  35  &  41  & 36  & 45   & 52   & 46    \\
\hline
POS  & 54 &  30  &  43  & 36  & 42   & 50   & 43  \\
\hline
LE PR.  & 6.2 & 3.4  &  4.9 & 4.2 & 5.8  & 5.8  & 5.8  \\
\hline
HE PR.  & 4.1 & 1.5  &  2.8 & 3.1 & 3.2  & 4.0  & 3.8   \\
\hline
\hline
Total     &  119 & 70  &  92  &  79 &  96 & 112  & 99  \\
\hline
\hline
ALB-1 & 22 & 20 & 17 & 18 & 19 & 21 & 18  \\
\hline
ALB-2 & 39 & 36 & 30 & 33 & 35 & 36 & 33  \\
\hline
\end{tabular}
\vspace{0.5 truecm}
\vspace{0.2 cm}

\caption{\label{tab:trig15-2}
{\bf Summary of Level-1.5  processing of cosmic gamma-ray events (*)}}

\begin{tabular}{l c c c c c c c c c}
\hline
  Photon     & R11G & DIS & RUD12 & RUD10 & VFORM & CGAP &
  TGAP  \\
  Class       &(TC) & (TC) & (TC) & (TC)& (TC)
 &(TC) & (TC) \\
  \hline
HE 0-10   & 40\% & 39\% & 39\% & 36\% & 35\% & 39\% & 38\% \\
\hline
HE 50-60  & 26\% & 25\% & 24\% & 21\% & 19\% & 26\% & 23\%  \\
\hline
LE 0-10   & 26\% & 26\% & 23\% & 20\% & 25\% & 22\% & 24\%  \\
\hline
LE 50-60  & 18\% & 17\% & 14\% & 12\% & 15\% & 16\% & 15\%  \\
\hline
\end{tabular}

\end{center}
\vspace{0.3truecm}
(*) Simulation results concerning only Tracker-Converted (TC) photons. Values in the table
represent the percentage of the number of simulated events that pass the different Level-1.5
trigger options respect to the total number
 of photons that theoretically could enter into the Tracker volume (\% of TRA\_TH)
\vspace{0.5truecm}
\end{table}
\vspace{0.2 cm}
\begin{table}[!h]
\begin{center}
\caption{\label{tab:trig15s} {\small \bf
Average GRID Background  Event Rates after Level-1$+$1.5
Processing}} \vskip .09in
\begin{tabular}{ l c c }
\hline
Background component    &       unocculted    &       half-occulted \\
                        &       GRID FOV      &       GRID FOV        \\
\hline
Charged particles      &  70 $\rm s^{-1}$     &     70 $\rm s^{-1}$ \\
Earth albedo-photons   &  20 $\rm s^{-1}$     &     40 $\rm s^{-1}$  \\
\hline
Total                  &  90 $\rm s^{-1}$     &    110 $\rm s^{-1}$ \\
\hline
\end{tabular}
\end{center}
\vspace{0.5truecm}
\end{table}


\end{document}